\documentclass[prl,letterpaper,twocolumn,showpacs,superscriptaddress]{revtex4}
\usepackage{graphicx,psfrag,amsmath,amssymb,amsfonts,bbm,latexsym,color,dcolumn}

\begin{document}

\title{Microscopic origin of critical current fluctuations in large, small
and ultra-small area Josephson junctions.}
\author{Lara Faoro}
\author{Lev B. Ioffe}
\affiliation{Department of Physics and Astronomy, Rutgers University, 136 Frelinghuysen
Rd, Piscataway 08854, New Jersey, USA}
\date{\today}

\begin{abstract}
We analyze data on the critical current and normal state resistance noise in
Josephson junctions and argue that the noise in the critical current is due
to a mechanism that is absent in the normal state. We estimate the noise
produced by conventional Two Level Systems (TLSs) in the insulating barrier and
find that it agrees both in magnitude and in temperature dependence with 
the resistance fluctuations in the normal state but it is not sufficient to
explain the critical current noise observed in large superconducting contacts. 
We propose a novel microscopic mechanism for the noise in the superconducting state 
in which the noise is due to electron
tunneling between weak Kondo states at subgap energies. We argue that the noise
produced by this mechanism gives temperature, area dependence and
intensity that agree with the data.     \newline
\end{abstract}

\pacs{85.25.Cp, 03.65.Yz,73.23.-b}
\maketitle




\textit{Introduction.$-$} The microscopic mechanism at the origin of
critical current fluctuations in Josephson junctions is a long standing open
problem. A phenomenological characterization of critical current noise began
with the experiments in large Niobium
junctions in the late $80$'s \cite{Wellstoodthesis}; in these experiments the critical current
fluctuations were inferred from the fluctuations in the current of
resistively shunted junctions biased by a low voltage that has no effect on
the superconductivity of the contacts themselves. It was found
that the noise power spectra display $1/f$ behavior for low frequencies ${%
f\lesssim 10 \text{Hz with }} T^{2}-$ temperature dependent intensity \cite%
{Wellstood2004} and that it also scales with the inverse area of the
junction with coefficient that depends, albeit weakly, on the
superconducting or insulating barrier material \cite{VanHarlingen2005}.

It is natural to attribute the critical current fluctuations to charges
that move between different localized states in the junction barrier \cite%
{RogersPRL85}. However, a more detailed comparison with the experiments reveals
serious problems with this picture.  In this model, each fluctuator is
similar to a glassy TLS \cite{Burin1995},
so one expects that the distance between the levels and the tunneling
barrier have a broad distribution on the atomic energy scale. Quantum
tunneling or thermal activation leads to the charge motion between two
states which changes the barrier height and thus leads to a noisy Random
Telegraph Signal (RTS) in the current through the junction 
\cite{RogersPRL84}.  A superposition of RTSs with a broad distribution of 
tunneling barriers leads to the $1/f$ noise power spectra \cite{Dutta1981}.  
The problem with this model is that TLSs and similar objects have a constant density of states
at low energies and this would give a linear $T-$dependence of the noise
power spectrum in contrast with the data. Alternatively, assuming that $T^{2}$
behavior comes from the linear density of states and extrapolating this
density of states to atomic scales one gets unphysically large values for
the density of these switches thus indicating the presence of some low
energy scale in the problem.

Recently, the interest in  critical current fluctuations was renewed
because of their importance to the superconducting qubit dephasing and a new
puzzle was added to the picture. The new experiments \cite{Eroms2006} studied the
fluctuations in the small area $(A\sim 0.1 \mu \text{m}^{2})$ Aluminium junctions, 
similar to the ones that are currently used for qubit
implementations in the flux qubit \cite{Chiorescu2003}, the phase qubit \cite%
{Martinis2002} and the quantronium \cite{Vion2002}. The critical current
noise was inferred from fluctuations in the \emph{normal} state resistance of
the junction. It has been observed that the noise power spectrum displays a $%
1/f$ behavior at low frequency but its temperature dependence is only linear
and the intensity of the noise is two order of magnitude lower than the
value reported for larger superconducting contacts.

In this Letter we argue that the problems with the microscopic model noted
above as well as the inconsitency between the old and the new data are all
removed if the main source of the critical current noise is electron
trapping in shallow subgap states that might be formed at the
superconductor-insulator boundary \cite{Faoro2005}.  The electron tunneling
between such traps contains two Fermi factors leading to the $T^{2}/T_{0}^2$
behavior with the energy scale $T_{0} \sim T_{c}$, thereby eliminating the difficulties with the conventional model noted above. This model is further
supported by the recent experiments performed in Single Electron Transistors
that show $T^{2}$ dependence of the low $1/f$ charge noise power spectra in very small superconducting contacts \cite{Astafiev2004}.
We shall argue that this mechanism dominates the critical current noise in
the superconductive regime but disappears in the normal state where these
states become Kondo resonances leaving only a weaker conventional TLSs
mechanism active that, at high
temperatures, produces linear $T-$dependence of the noise power in agreement
with the data \cite{Eroms2006}. An important feature of the new mechanism
is that the large number of subgap states is partially compensated by
their small weight (a vestige of their Kondo resonance origin) \cite%
{Faoro2005}. Physically, it means that these states generate a featureless $1/f$
noise even in the smallest contacts. In contrast, in normal leads, where the noise is 
due to TLSs in the insulating barrier, our estimates show that the number
of active fluctuators, especially in ultrasmall contacts, is low, in agreement with the data \cite{RogersIEEE85}.

Below we discuss the details of the data and estimates of the noise produced
in new and conventional microscopic models. We begin with the data.

\emph{Data.$-$} The directly measured low frequency noise (${f\leq 10\text{Hz}}$%
) of the critical current turns out to be almost universal. The original
experiments measured it in large contacts (area, $A$, varying between ${%
10-100\mu \text{m}^{2}}$) made of ${\text{Nb-NbO}_{\text{x}}\text{-PbIn}}$ 
\cite{Wellstoodthesis,Wellstood2004}. In these experiments the
junctions were biased by a very low voltage ${V\approx 1-5\mu V}$ 
that does not affect the superconductivity of the 
contacts (i.e. $V{\ll 2 \Delta /e}$) \cite{Cri1}. In
these conditions, the noise power spectra are described by: 
\begin{equation}
\frac{S_{I_{0}}(\omega )}{I_{0}^{2}}=\gamma \frac{A_{0}}{T_{0}^{2}}\frac{%
T^{2}}{A\omega }\;  \label{ciao} 
\end{equation}%
in a wide temperature range ${0.09 \text{K} \leq T\leq 4.2\text{K}}$. Here $I_0$ is the critical current, $A_{0}$ and $T_{0}$ are the area and temperature scale which are
conventionally set to $A_{0}=1\mu \text{m}^{2}$ and $T_{0}=4.2 \text{K}$. Remarkably, the
dimensionless proportionality coefficient, $\gamma $, is not sensitive to
the details of the junction preparation and it is roughly (within a factor $3)$
\textquotedblleft universal \textquotedblright: ${\gamma \approx
1.44 \cdot10^{-10}}$ \cite{VanHarlingen2005}. The noise observed in these
experiments was usually featureless, evidently coming from many fluctuators,
but in rare cases and at relatively high temperatures, one observes also a
switching process between two well defined current values that
disappears when temperature decreases below $T<1$K. Notice that
the quadratic growth of the noise with temperature that starts at ${T=0.1\text{K}}$ implies that 
the number of fluctuators at atomic energies would be at least $10^{10}$ larger
than their number at low temperatures and thus would exceed the number of atoms
in $10\mu \text{m}^{2}$ contact ($N\lesssim 10^{9}$), providing the evidence of additional energy scale.

In a different set of experiments \cite{Eroms2006}, the critical current of small junctions
was not directly measured but extracted from their normal state resistance
fluctuations. To perform measurements at very low temperatures, the ${
\text{Al-AlO}_{x}\text{-Al}}$ junctions of ${A\sim 0.1\mu \text{m}^{2}}$ area were
subjected to the magnetic field above ${0.1\text{T}}$ that
suppressed the superconductivity in the contacts. It was found that for ${0.3%
\text{K}\leq T\leq 5\text{K}}$ the noise power spectrum is well described by
a linear $T$-dependence:
\begin{equation}
\frac{S_{R_{n}}(\omega )}{R_{n}^{2}}\propto \frac{T}{A\omega }\;
\label{Mooij}
\end{equation}%
with the intensity that is two order of magnitudes lower than that predicted
by Eq.~(\ref{ciao}). Moreover, for temperatures below $0.8\text{K}$, it has been
observed that the low frequency noise was due to few
individual strong fluctuators. These observations are in a good agreement
with the conventional TLSs picture of the noise origin. Indeed, in a typical
structural glass one expects a constant density of states ${\nu \sim
10^{20}cm^{-3}e\text{V}^{-1}}$ that leads to the linear temperature dependence of
the noise and to the total number of the thermally excited TLSs, ${{\cal N}_{\text{TLSs}}(T=1\text{K})
\sim 10}$, in ${1\mu \text{m}^{2}}$ contact. These data are also in a
qualitative agreement with the older measurements of ultrasmall (${A\leq
0.05\mu \text{m}^{2}\text{) Nb-Nb}_{2}\text{O}_{5}\text{-PbIn}}$ junctions
performed in the temperature range ${1\text{K}\leq T\leq 300\text{K}}$  \cite{RogersIEEE85}. In
these measurements, the contacts either were kept normal due to a high temperature
or they were very far from equilibrium due to a high voltage bias ${V>10 \text{mV}\gg 2\Delta
/e}$;  the observed low frequency $1/f$ noise
was dominated by few fluctuators even at $T<100\text{K}$.

The detailed quantitative comparison between different data and microscopic
estimates is made difficult by the possibility of the inhomogeneous current
distribution through the insulating barrier. For instance, the newer data \cite{Eroms2006} report the relative variation of the resistivity due to one
fluctuator corresponding to the change in the \emph{effective} area of
the junction by $\delta A_{eff}=1\text{nm}^{2}$  while for the older data \cite%
{RogersIEEE85} this value is  $\delta A_{eff}\sim 10 \text{nm}^{2}$. This
difference explains why the noise in new experiments got dominated by a few
fluctuators only below $1\text{K}$ in contrast to \cite{RogersPRL84}
where single fluctuators were resolved below $100 \text{K}$. Indeed, it is difficult
to imagine microscopic processes that affect large areas $\sim 10 \text{nm}^{2}$; a
more plausible explanation is that the conductance is dominated by 
relative small areas or channels in this system \cite{Buchanan02} which
decreases further the number of fluctuators (Figure 1). This conjecture is
supported by the recent analysis  \cite{Dorneles03} that found that the
conductive area is $\kappa \sim 10^{-5}$ of the total. Most likely, however,
this value depends strongly on  the material preparation. 

\begin{figure}[tbph]
\centering
\includegraphics[width=1.8 in]{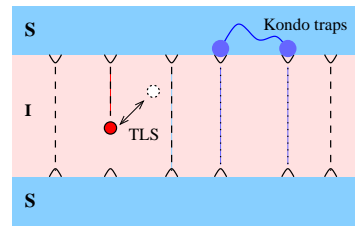}
\caption{Sketches of the microscopic mechanisms of decoherence that might be responsible for critical current noise.}
\end{figure}

\emph{Microscopic mechanisms of decoherence.$-$} All mechanisms of the noise
assume that it comes from the fluctuations in the state of the  insulator
that affect the tunneling barrier and thus the critical current. If these
fluctuations are not affected by the state of the metal (superconductor or
normal) they would equally affect the critical current $I_{0}$ and the
normal state resistance $R_{n}$ according to the Ambegaokar-Baratoff formula: 
${I_{0}=\frac{\pi }{2eR_{n}}\Delta(T)\tanh \left( \frac{\Delta (T)}{K_{B}T}\right)}$ \cite{Ambegaokar63}.  One can argue that these fluctuations might be due to fluctutating  trapped electrons or randomly moving TLSs inside the barrier.

Low energy electrons trapped in the vicinity of the
normal metal can tunnel into the metal; this process converts the sharp
state into a short-living resonance. In superconductor, however, the low energy states
disappear and the life time of these states increases. Notice that  a large potential 
difference between the contacts provides electron states below the gap and thus makes the
system effectively normal. In a superconductor a localized low energy
electron can change its state only by tunneling into another subgap state
via Andreev process \cite{Faoro2004}. If the density of states for these
subgap states is constant, the number of thermally active electrons and
their possible final states both scale with temperature leading to $T^{2}$
behavior of the noise in agreement with the data. However, similarly to TLSs,
the density of states for such traps is too little to account for the number
of fluctuators observed experimentally.

We believe that the physical mechanism responsible for a dramatic increase
in the density of states for the localized low energy electrons is the
formation of Kondo-like resonances due to a large Coulomb repulsion between
the electrons in the same trap \cite{Faoro2005}.  In this picture, the characteristic
energy scale for these resonances is given by the Kondo temperature $T_{K}$
which depends exponentially on the bare level width $\Gamma $ and the bare
level position $\epsilon _{0}$: ${T_{K}\propto \exp (-\pi \epsilon
_{0}/2\Gamma )}$. The natural assumption that $\Gamma $ is
distributed in a broad range leads to ${P(T_{K})\propto 1/T_{K}}$
distribution of Kondo temperatures in the normal state. In the
superconductor the resonances having ${T_{K}^{\ast }\approx 0.3\Delta }$
become localized low energy levels with a constant surface density of states 
${\nu (\epsilon )=\rho _{2D}/({\mathfrak{P}}T_{K}^{\ast })}$ where $\rho
_{2D}$ is the bare surface density of traps (${\rho _{2D}{\approx
10^{14}/\text{cm}^{2}}}$) and ${{\mathfrak{P}}\approx 1}$. The increase in the
density of states of these objects is partially offset by the small weight
of the Kondo resonance ${w=T_{K}/\epsilon _{0}}$ that translates into the
small weight of the formed localized state. 

The noise at frequency $f$ is due to the electron tunneling between two
traps with rate ${\tau ^{-1}\sim f}$. The exact calculation of the rate for
two specific traps is a complicated problem but it is sufficient for our
purposes to note that the latter is an exponential function of the bare parameters and
thus it is characterized by ${dP(\tau )=d\tau /\tau }$ distribution function.
Indeed, in superconductors the tunneling amplitude falls exponentially with
distance: ${\widetilde{t}\propto {e^{-r/\xi }}}$ where $\xi $ is the coherence
length of the superconductor. Moreover, the tunneling between Kondo resonances is
additionally suppressed by the small factor $w$ and finally, the electron tunneling between the
states with different energies is also accompanied by the thermal phonon
emission or absorption but the rate of these processes is much faster than
the frequencies of interest ($f<1kHz$) so the main dependence comes from the
exponential factors mentioned above.  

In order to estimate the critical current fluctuations induced by these
tunneling processes we need to know the effect induced by a single electron
trapped in a deep state. We shall describe it by the change in the effective
area of the junction ${\delta \widetilde{A}_{eff}}$ (i.e. $\displaystyle{%
\delta I_{0}=\frac{I_{0}}{A}\delta }\widetilde{A}_{eff}$). For the
quantitative estimates we shall assume that ${\delta \widetilde{A}_{eff}\sim
\delta A_{eff}}$. The tunneling processes allow tunneling between
the traps within energy $T$ from the Fermi surface located at distances 
${r\lesssim \xi}$ from each other. Combining all factors we find that the
noise spectrum generated by independent relaxational processes between the
traps reads: 
\begin{eqnarray}
S_{I_{0}}(\omega ) &\approx &I_{0}^{2}\frac{\delta \widetilde{A}_{eff}^{2}}{%
A^{2}}w^{2}T^{2}{A\xi ^{2}\nu }^{2}{(\epsilon )}\int dP(\tau )\frac{\tau }{%
1+\omega ^{2}\tau ^{2}}\;  \notag \\
&=&c\frac{I_{0}^{2}}{A}\left( w\rho _{2D}\delta \widetilde{A}_{eff}\xi
\right) ^{2}\left( \frac{T}{T_{K}^{\ast }}\right) ^{2}\frac{1}{\omega }
\label{Kondo}
\end{eqnarray}%
where ${c \sim O(1)}$. This noise displays $T^{2}$ dependence and
inverse proportionality to the area of the contacts in agreement with the data
observed in large contacts. To estimate the intensity of the noise we assume that the 
distance between electron traps is roughly of the same order as the
distance between TLSs in the bulk or (between typical surface defects)
leading to $\rho _{2D}\approx 10^{13}{cm^{-2}}.$ Using the Nb parameters:
${\Delta _{\text{Nb}}=16.33\text{K}}$, ${\xi _{\text{Nb}}\approx 40\text{nm}}$, 
$\delta A_{eff}\sim 5 \text{nm}^{2}$, $w=10^{-3}$, we estimate the
dimensionless parameter $\gamma$ controlling  the noise intensity at
$T=4.2\text{K}$, obtaining $\gamma \approx 5\cdot10^{-10}$ which is in excellent agreement
with the measured value (given the crude nature of the estimates). 

Alternatively, critical current noise might be due to fluctuating TLSs in the insulating barrier \cite{Burin1995}.  Most likely, TLSs correspond to atoms that can tunnel between two positions in the amorphous material. Two different states can be distinguished by the different values
of the dipole moment. Thus, TLSs deep inside the insulator interact only
weakly with electrons in the metal and are not affected by the
superconductivity. Each quantum TLS is described by the Hamiltonian ${%
H_{0}=\varepsilon \sigma _{z}+t\sigma _{x}}$, where $\varepsilon $ is the
energy difference between the two minima and ${t=\hbar \omega _{0}e^{-S}}$
is the tunneling amplitude between them. Here $\omega _{0}$ is of the order
of the frequency vibrations of the particle in the potential wells and ${S \gg 1}$
is the tunneling action. In bulk amorphous materials $\epsilon $ and $S$ are
distributed in a broad range leading to the distribution function 
${P(\epsilon ,t)=\nu /t}$. Generally, each TLS has states with energies ${E=\pm 
\sqrt{\varepsilon ^{2}+t^{2}}}$ that are occupied with thermal probability
factors. At not-too-low temperatures ($T\gtrsim 0.2\text{K}$) the TLS relaxation
process is dominated by phonon emission or absorption that leads to the
relaxation rate of the occupancy number ${\tau _{ph}^{-1}=aEt^{2}\coth \beta E}$ where the 
coefficient $a$ depends on the details of the TLS-phonon
interaction \cite{KoganBook}. The dominant contribution to low frequency
noise comes from thermally excited TLSs with low tunneling amplitudes $t\ll
\epsilon \sim T$. These TLSs are characterized by the distribution function ${P(E,\tau )=\nu /2\tau}$. Each individual TLS contributes:   
${S_{I_{0}}^{(1)}=\frac{\delta I_{0}^{2}}{\cosh ^{2}\beta E}\frac{\tau }{%
1+\omega ^{2}\tau ^{2}}\;}$  
to the current-current correlator.  By assuming as before that each TLS affects
the critical current by ${\delta I_{0}=(\delta A/A)I_{0}}$ and by averaging over 
the TLSs distribution function we find that the total noise power spectrum
originating from the uniform insulating barrier of thickness $d$ reads:  
\begin{equation}
S_{I_{0}}(\omega )\simeq \kappa \frac{I_{0}^{2}}{A}\delta A^{2}\nu d\frac{T}{%
\omega }.\;  \label{SpeTLS}
\end{equation}
As expected this noise is proportional to $T$ and $1/A$.  Non uniform
distribution of the current discussed above decreases the number of the
effective TLSs because each TLS affects only the current that flows in a
close vicinity of a moving atom. We describe
this suppression by the dimensionless coefficient $\kappa $; very crudely we
can estimate ${\kappa \sim \delta A_{0}/\delta A}$ where ${\delta
A_{0}=0.1\text{nm}^{2}}$.  For small Al contacts,  $A=1\mu \text{m}^{2}$, at $T=4.2\text{K}$
we get ${S_{I_0}/I_0^{2} \approx 3 \cdot 10^{-12}}$, which is in agreement with the direct data on the normal
state resistance fluctuations \cite{Eroms2006} . Notice that a relative
small value of the effective area $\delta A$ for these contacts implies that
the current is relatively homogeneous; one expects smaller values for the
contacts studied earlier \cite{RogersIEEE85}  for which similar
estimates gives a larger result ${S_{I_0}/I_0^{2} \approx 3 \cdot 10^{-11}}$ which is however
still smaller that the noise observed directly in the superconducting state.

As a final remark, let us notice that the electron trap model for the noise in the superconducting 
state leads to an interesting prediction. Namely, the process of tunneling between these traps
should also contribute to the high frequency dissipation leading to
\emph{ohmic} behaviour of the spectrum: ${S_{I_{0}}(\omega )=2e}^{2}G{\omega }$. 
The proportionality constant, ${G}$, in this formula has a meaning of
the effective conductance of the contact due to the trapped electrons, i.e. 
${G\approx \gamma \frac{A_{0}}{A}\left( \frac{\pi \Delta }{2e^{2}R_{n}T_{0}}
\right) ^{2}}$, and it could be
in principle extracted from measurements similar to the ones performed by 
Astafiev et al. \cite{Astafiev2004} for the charge noise.  Here, however,
such experiments are made difficult by the noise coming from the shunt resistors and the 
residual resistance of the contact itself due to charge fluctuators. 

\textit{Conclusions. $-$} We identify a new microscopic mechanism responsible
for the noise in the superconducting contacts. This mechanism contributes to
the low frequency noise only in the superconducting state and it is due to the
appearance of subgap states localized near the superconductor-insulator
boundary.  The resulting noise has homogeneous spectrum, $T^{2}$
dependence and it is inversely proportional to the surface area $%
A$ of the junction in agreement with the experiments. The estimates of its
intensity are in a good agreement with experimental values. This mechanism
disappears when the superconductivity of the contacts is suppressed. In the
normal state the noise is generated by the thermally excited TLSs. In this case, the noise 
power spectra of the critical current fluctuations show a linear $T$ dependence, inverse
proportionality to the area, with intensity of the noise lower than the one
generated by  the superconducting state in agreement with experiments. The
mechanism that dominates the critical current noise should also dominate the charge noise.
It would be important to confirm it directly by measuring
the charge and critical current/resistance noise on the same (or at least
similar) samples in both superconducting and normal states. \acknowledgments %
We thank O. Astafiev, J. Clarke, M. Gershenson, D. J. Van Harlingen, J.
Martinis, R. W. Simmonds and F. Wellstood for useful discussions. This work
was supported by the National Security Agency (NSA) under Army Research
Office (ARO) contract number W911NF-06-1-0208 and NSF ECS 0608842. 

\vspace*{-2mm}

\end{document}